\documentclass[12pt]{article}
\usepackage{epsfig}
\usepackage{amsmath}
\usepackage{latexsym}
\title{Quantum Corrections to the Reissner-Nordstr\"{o}m and
Kerr-Newman Metrics: Spin 1}
\author{Barry R. Holstein\\
Department of Physics-LGRT\\
University of Massachusetts\\
Amherst, MA  01003}
\begin{document}
\begin{titlepage}
\maketitle
\begin{abstract}
A previous evaluation of one-photon loop corrections to the
energy-momentum tensor has been extended to particles with unit
spin and speculations are presented concerning general properties
of such forms.
\end{abstract}

\end{titlepage}

\section{Introduction}

In an earlier letter, we described calculations of the one-loop
radiative corrections to the energy-momentum tensor of a charged
spinless or spin 1/2 particle of mass $m$ and focused on the
nonanalytic component of such results\cite{gar}.   This is because
such nonanalytic pieces involve terms with singularities at small
momentum transfer q which, when Fourier-transformed, yield---via
the Einstein equations---large distance corrections to the metric
tensor. In particular, for both the spinless and spin 1/2 field
case, the diagonal component of the metric was shown to be
modified from its simple Schwarzschild form via terms which
account for the feature that the particle is
charged---specifically, in harmonic gauge the resulting metric has
the form
\begin{eqnarray}
g_{00}&=&1-{2Gm\over r}+{G\alpha\over r^2}-{8 G\hbar\over 3\pi
mr^3}+\ldots\nonumber\\
g_{ij}&=&-\delta_{ij}-\delta_{ij}{2Gm\over
r}+{G\alpha}{r_ir_j\over r^4}+{4\over 3\pi}{G\alpha\hbar\over
mr^3}\left({r_ir_j\over r^2}-\delta_{ij}\right)\nonumber\\
&-&{4\over 3\pi} {G\alpha\hbar(1-\log\mu r)\over
mr^3}\left(\delta_{ij}-3{r_ir_j\over r^2}\right)
\end{eqnarray}
where $G$ is the gravitational constant and $\alpha=e^2/4\pi$ is
the fine structure constant.  (Note that the dependence on the
arbitrary scale factor $\mu$ can be removed by a coordinate
transformation.) The classical---$\hbar$-independent---pieces of
these $\alpha$-dependent modifications are well known and can be
found by expanding the familiar Reissner-Nordstr\"{o}m metric,
describing spacetime around a charged, massive object\cite{rnm}.
On the other hand, the calculation also yields quantum
mechanical---$\hbar$-dependent---pieces which are new and whose
origin can be understood qualitatively as arising from
zitterbewegung\cite{gar}.

\begin{figure}
\begin{center}
\epsfig{file=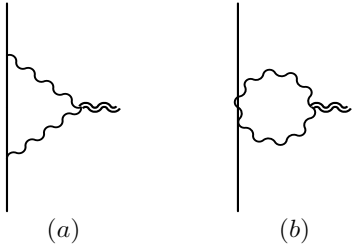,height=4cm,width=7cm}
 \caption{Feynman diagrams having nonanalytic components.  Here
 the single wiggly lines represent photons while the double
 wiggly line indicates coupling to a graviton.}
 \end{center}
\end{figure}

In the case of a spin 1/2 system there exists, in addition to the
above, a nonvanishing off-diagonal component of the metric, whose
radiative corrected form, in harmonic gauge, is
\begin{equation}
g_{0i}=(\vec{S}\times\vec{r})_i\left({2G\over r^3}-{G\alpha\over
mr^4} +{2G\alpha\hbar\over \pi m^2r^5}+\ldots\right)
\end{equation}
Here the classical component of this modification can be found by
expanding the Kerr-Newman metric\cite{knm}, which describes
spacetime in the vicinity of a charged, massive, and spinning
particle, and again there exist quantum corrections due to
zitterbewegung\cite{gar}.

Based on the feature that the diagonal components were found to
have an identical form for both spin 0 and 1/2, it is tempting to
speculate that the leading diagonal piece of the metric about a
charged particle has a universal form---independent of spin.
Whether the same is true for the leading
off-diagonal---spin-dependent---component cannot be determined
from a single calculation, but it is reasonable to speculate that
this is also the case. However, whether these assertions are
generally valid can be found only by additional calculation, which
is the purpose of the present note, wherein we evaluate the
nonanalytic piece of the radiatively-corrected energy-momentum
tensor for a charged spin 1 particle---the $W^\pm$ boson---and
assess the correctness of this proposal. In the next section then
we briefly review the results of the previous paper, and follow
with a discussion wherein the calculations are extended to the
spin 1 system. Results are summarized in a brief concluding
section.

\section{Lightning Review}

Since it important to the remainder of this note, we first present
a concise review of the results obtained in our previous
paper\cite{gar}.  In the case of a spin 0 system, the general form
of the energy-momentum tensor is
\begin{equation}
<p_2|T_{\mu\nu}(x)|p_1>_{S=0}={e^{i(p_2-p_1)\cdot x}\over
\sqrt{4E_2E_1}}\left[2P_\mu P_\nu F_1^{(S=0)}(q^2))+(q_\mu
q_\nu-q^2\eta_{\mu\nu})F_2^{(S=0)}(q^2)\right]
\end{equation}
where $P={1\over 2}(p_1+p_2)$ is the average momentum, while
$q=p_1-p_2$ is the momentum transfer.  The tree level values for
these form factors are
\begin{equation}
F_{1,tree}^{(S=0)}=1\qquad F_{2,tree}^{(S=0)}=-{1\over 2}
\end{equation}
while the leading nonanalytic loop corrections from Figure 1a and
Figure 1b were determined to be
\begin{eqnarray}
F_{1,loop}^{(S=0)}(q^2)&=&{\alpha\over 16\pi}{q^2\over
m^2}(8L+3S) \nonumber\\
F_{2,loop}^{(S=0)}(q^2)&=&{\alpha\over 24\pi}(8L+3S)
\end{eqnarray}
where we have defined
$$L=\log ({-q^2\over m^2})\quad{\rm and}\quad S=\pi^2\sqrt{m^2\over
-q^2}.$$ Such nonanalytic forms, which are singular in the small-q
limit, are present due to the presence of {\it two} massless
propagators in the Feynman diagrams\cite{doh} and can arise even
in electromagnetic diagrams when this situation exists\cite{emf}.
Upon Fourier-transforming, the piece proportional to $S$ is found
to give classical ($\hbar$-independent) behavior while the term
involving $L$ is found to yield quantum mechanical
($\hbar$-dependent) corrections.\footnote{The at first surprising
feature that a loop calculation can yield classical physics is
explained in ref. \cite{doh}} The feature that the form factor
$F_1^{(S=0)}(q^2=0)$ remains unity even when electromagnetic
corrections are included arises from the stricture of
energy-momentum conservation\cite{gar}. There exists no
restriction on $F_2^{(S=0)}(q^2=0)$.

In the case of spin 1/2 there exists an additional form
factor---$F_3^{(S={1\over 2})}(q^2)$---associated with the
existence of spin---
\begin{eqnarray}
<p_2|T_{\mu\nu}(x)|p_1>_{S={1\over 2}}&=&{e^{i(p_2-p_1)\cdot
x}\over \sqrt{E_1E_2}} \bar{u}(p_2)\left[P_\mu P_\nu
F_1^{(S={1\over
2})}(q^2)\right.\nonumber\\
&+&\left.{1\over 2}(q_\mu q_\nu -q^2\eta_{\mu\nu})F_2^{(S={1\over
2
)}}(q^2)\right.\nonumber\\
&-&\left.\left({i\over 4}\sigma_{\mu\lambda}q^\lambda
P_\nu+{i\over 4} \sigma_{\nu\lambda}q^\lambda
P_\mu\right)F_3^{(S={1\over 2})}(q^2)\right]u(p_1)\label{eq:bb}
\end{eqnarray}
The tree level values for these form factors are
\begin{equation}
F_{1,tree}^{(S={1\over 2})}=F_{3,tree}^{(S={1\over 2})}=1\qquad
F_{2,tree}^{(S={1\over 2})}=0
\end{equation}
while the nonanalytic loop corrections from Figure 1a and Figure
1b were found to be
\begin{eqnarray}
F_{1,loop}^{(S={1\over 2})}(q^2)&=&{\alpha\over 16\pi}{q^2\over
m^2}(8L+3S) \nonumber\\
F_{2,loop}^{(S={1\over 2})}(q^2)&=&{\alpha\over
24\pi}(8L+3S)\nonumber\\
F_{3,loop}^{(S={1\over 2})}(q^2)&=&{\alpha\over 24\pi}{q^2\over
m^2}(4L+3S)\label{eq:rw}
\end{eqnarray}
In this case both $F_1^{(S={1\over 2})}(q^2=0)$ {\it and}
$F_3^{(S={1\over 2})}(q^2=0)$ retain their value of unity even in
the presence of electromagnetic corrections.  That this must be
true for $F_1^{(S={1\over 2})}(q^2=0)$ follows from
energy-momentum conservation, as before, while the
nonrenormalization of $F_3^{(S={1\over 2})}(q^2=0)$ is required by
angular-momentum conservation\cite{amc} and asserts that an {\it
anomalous} gravitomagnetic moment is forbidden.  The universality
of these radiatively corrected forms is suggested by the result
\begin{equation}
F_{1,loop}^{(S=0)}(q^2)=F_{1,loop}^{(S={1\over 2})}(q^2)\quad{\rm
and} \quad F_{2,loop}^{(S=0)}(q^2)=F_{2,loop}^{(S={1\over
2})}(q^2)
\end{equation}
but, of course, the spin-dependent gravitomagnetic form factor
$F_3^{(S={1\over 2})}(q^2)$ has no analog in the spin 0 sector.

The connection with the metric tensor described in the
introduction arises when these results for the energy-momentum
tensor are combined with the (linearized) Einstein
equation\cite{eeq}
\begin{equation}
\Box h_{\mu\nu}=-16\pi G\left(T_{\mu\nu}-{1\over 2} \eta_{\mu\nu}
T\right)
\end{equation}
where we have defined
\begin{equation}
g_{\mu\nu}=\eta_{\mu\nu}+h_{\mu\nu}
\end{equation}
and
\begin{equation}
T\equiv {\rm Tr}\,T_{\mu\nu}
\end{equation}

 Taking Fourier transforms, we find---for both spin 0 and spin
1/2---the diagonal components
\begin{eqnarray}
h_{00}(\vec{r})&=&-16\pi G\int{d^3k\over
(2\pi)^3}e^{i\vec{k}\cdot\vec{r}}{1\over \vec{k}^2}\left({m\over
2} -{\alpha\pi|\vec{k}|\over 8}-{\alpha\vec{k}^2\over 3\pi
m}\log{\vec{k}^2\over m^2}\right)\nonumber\\
h_{ij}(\vec{r})&=&-16\pi G\int{d^3k\over
(2\pi)^3}e^{i\vec{k}\cdot\vec{r}}{1\over \vec{k}^2}\left({m\over
2}\delta_{ij} +({\alpha\pi\over 16|\vec{k}|}+{\alpha\over 6\pi
m}\log{\vec{k}^2\over m^2})(k_ik_j-\delta_{ij}\vec{k}^2)\right)\nonumber\\
\quad
\end{eqnarray}
while from the spin 1/2 gravitomagnetic form factor we find the
off-diagonal term
\begin{equation}
h_{0i}(\vec{r})=-16\pi Gi\int{d^3k\over (2\pi)^3}{1\over
\vec{k}^2} \left({1\over 2}-{\alpha\pi|\vec{k}|\over 16
m}-{\alpha\vec{k}^2\over 12\pi m^2}\log{\vec{k}^2\over m^2}
\right)(\vec{S}\times\vec{k})_i
\end{equation}
Evaluating the various Fourier transforms, we find the results
quoted in the introduction\cite{fta}.

The purpose of the present note is to study how these results
generalize to the case of higher spin.  Specifically, we shall
below examine the radiative corrections to the energy-momentum
tensor of a charged spin 1 system.

\section{Spin 1}

A neutral spin 1 field $\phi_\mu(x)$ having mass $m$ is described
by the Proca Lagrangian, which is of the form\cite{pro}
\begin{equation}
{\cal L}(x)=-{1\over 4}U_{\mu\nu}(x)U^{\mu\nu}(x)+{1\over
2}m^2\phi_\mu(x)\phi^\mu(x)\label{eqn:la}
\end{equation}
where
\begin{equation}
U_{\mu\nu}(x)=i\partial_\mu \phi_\nu(x)-i\partial_\nu\phi_\mu(x)
\end{equation}
is the spin 1 field tensor.  If the particle has charge $e$, we
can generate a gauge-invariant form of Eq. \ref{eqn:la} by use of
the well-known minimal substitution\cite{msu}---defining
\begin{equation}
\pi_\mu=i\partial_\mu-eA_\mu(x)
\end{equation}
and
\begin{equation}
U_{\mu\nu}(x)=\pi_\mu\phi_\nu(x)-\pi_\nu \phi_\mu(x)
\end{equation}
the charged Proca Lagrangian density becomes
\begin{equation}
{\cal L}(x)=-{1\over 2}U_{\mu\nu}^\dagger(x)
U^{\mu\nu}(x)+m^2\phi_\mu^\dagger(x)\phi^\mu(x)
\end{equation}
Introducing the left-right derivative
\begin{equation}
D(x)\overleftrightarrow{\nabla} F(x)\equiv D(x)\nabla F(x)-(\nabla
D(x)) F(x)
\end{equation}
the single-photon component of the interaction can be written as
\begin{equation}
{\cal L}_{int}(x)=ieA^\mu(x)
\phi^{\alpha\dagger}(x)[\eta_{\alpha\beta}\overleftrightarrow{\nabla}_\mu
-\eta_{\beta\mu}\nabla_\alpha]\phi^\beta(x)+\eta_{\alpha\mu}(\nabla_\beta
\phi^{\alpha\dagger}(x))\phi^\beta(x)
\end{equation}
so that the on-shell matrix element of the electromagnetic current
becomes
\begin{equation}{1\over
\sqrt{4E_fE_i}}<p_f,\epsilon_B|j_\mu|p_i,\epsilon_A>=-{e\over
\sqrt{4E_fE_i}}\left[2P_\mu\epsilon_B^*\cdot\epsilon_A-
\epsilon_{A\mu}\epsilon_B^*\cdot
q+\epsilon_{B\mu}^*\epsilon_A\cdot q\right]\label{eq:br}
\end{equation}
where we have used the property
$p_f\cdot\epsilon_B^*=p_i\cdot\epsilon_A=0$ for the Proca
polarization vectors.  If we now look at the spatial piece of this
term we find
\begin{equation}
{1\over
\sqrt{4E_fE_i}}<p_f,\epsilon_B|\vec{\epsilon_\gamma}\cdot\vec{j}
|p_i,\epsilon_A>\simeq {e\over
2m}\vec{\epsilon}_\gamma\times\vec{q}\cdot\hat{\epsilon}_B^*\times\hat{\epsilon_A}
={e\over 2m}<1,m_f|\vec{S}|1,m_i>\cdot\vec{B}
\end{equation}
where we have used the result that in the Breit frame for a
nonrelativistically moving particle
\begin{equation}
i\hat{\epsilon}_B^*\times\hat{\epsilon}_A=<1,m_f|\vec{S}|1,m_i>
\end{equation}
which we recognize as representing a magnetic moment interaction
with g=1. On the other hand if we take the time component of Eq.
\ref{eq:br}, we find, again in the Breit frame and a
nonrelativistically moving system
\begin{equation}
{1\over \sqrt{4E_fE_i}}<p_f,\epsilon_B|\epsilon_{0\gamma }j_0
|p_i,\epsilon_A>\simeq -e\epsilon_{0\gamma}\left[
\epsilon_B^*\cdot\epsilon_A +{1\over 2m
}(\epsilon_{A0}\hat{\epsilon}_B^*\cdot\vec{q}-
\epsilon_{B0}^*\hat{\epsilon}_A\cdot\vec{q})\right]
\end{equation}
Using
\begin{eqnarray}
\epsilon_A^0\simeq{1\over 2m}\hat{\epsilon}_A\cdot\vec{q},\qquad
\epsilon_B^0\simeq -{1\over
2m}\hat{\epsilon}_B^*\cdot\vec{q}\nonumber\\
\epsilon_B^*\cdot\epsilon_A\simeq
-\hat{\epsilon}_B^*\cdot\hat{\epsilon}_A-{1\over
2m^2}\hat{\epsilon}_B^*\cdot\vec{q}\hat{\epsilon}_A\cdot\vec{q}
\end{eqnarray}
we observe that
\begin{equation}
{1\over \sqrt{4E_fE_i}}<p_f,\epsilon_B|\epsilon_{0\gamma }j_0
|p_i,\epsilon_A>\simeq e\epsilon_{0\gamma}
\hat{\epsilon}_B^*\cdot\hat{\epsilon}_A
\end{equation}
which is the expected electric monopole term---any electric
quadrupole contributions have cancelled\cite{yob}.  Overall then,
Eq. \ref{eq:br} corresponds to a simple E0 interaction with the
charge accompanied by an M1 interaction with g-factor unity, which
is consistent with the speculation by Belinfante that for a
particle of spin $S$, $g=1/S$\cite{bfc}.

Despite this suggestively simple result, however, Eq. \ref{eqn:la}
does {\it not} correctly describe the interaction of the charged
$W$-boson field, due to the feature that the $W^\pm$ are
components of an SU(2) vector field\cite{sut}. The proper Proca
Lagrangian has the form
\begin{equation}
{\cal L}(x)=-{1\over
4}\vec{U}_{\mu\nu}^\dagger(x)\cdot\vec{U}^{\mu\nu}(x)+{1\over
2}m_W^2\vec{\phi}_\mu(x)\cdot\vec{\phi}^\mu(x)\label{eqn:ld}
\end{equation}
where the field tensor $\vec{U}_{\mu\nu}(x)$ contains an
additional term on account of gauge invariance
\begin{equation}
\vec{U}_{\mu\nu}(x)=\pi_\mu\vec{U}_\nu(x)-\pi_\nu\vec{U}_\mu(x)
-ig\vec{U}_\mu(x)\times\vec{U}_\nu(x)
\end{equation}
with $g$ being the SU(2) electroweak coupling coupling constant.
The Lagrange density Eq. \ref{eqn:ld} then contains the piece
\begin{equation}
{\cal
L}_{int}(x)=-gW^{0\mu\nu}(x)(W_\mu^{+\dagger}(x)W_\nu^+(x)-W_\mu^{-\dagger}(x)W^-_\mu)(x)
\end{equation}
among (many) others.  However, in the standard model the neutral
member of the W-triplet is a linear combination of $Z^0$ and
photon fields\cite{smb}---
\begin{equation}
W_\mu^0=\cos\theta_WZ_\mu^0+\sin\theta_WA_\mu
\end{equation}
and, since $g\sin\theta_W=e$, we have a term in the interaction
Lagrangian
\begin{equation}
{\cal
L}_{int}^{(1)}(x)=-eF^{\mu\nu}(x)(W_\mu^{+\dagger}(x)W_\nu^+(x)-W_\mu^{-\dagger}(x)W^-_\mu(x))
\end{equation}
which represents an additional interaction that must be appended
to the convention Proca result.  In the Breit frame and for a
nonrelativistically moving system we have
\begin{equation}
{1\over
\sqrt{4E_fE_i}}<p_f,\epsilon_B|\vec{\epsilon_\gamma}\cdot\vec{j}^{(1)}
|p_i,\epsilon_A>\simeq {e\over
2m_W}\vec{\epsilon}_\gamma\times\vec{q}\cdot\hat{\epsilon}_B^*\times\hat{\epsilon_A}
={e\over 2m_W}<1,m_f|\vec{S}|1,m_i>\cdot\vec{B}\label{eq:j1}
\end{equation}
and
\begin{equation}
{1\over \sqrt{4E_fE_i}}<p_f,\epsilon_B|j_0^{(1)}
|p_i,\epsilon_A>\simeq -e{1\over 2m_W
}(\epsilon_A^0\hat{\epsilon}_B^*\cdot\vec{q}-
\epsilon_{B0}^*\hat{\epsilon}_A\cdot\vec{q})=-{e\over
2m_W^2}\hat{\epsilon}_B^*\cdot\vec{q}\hat{\epsilon}_A\cdot\vec{q}\label{eq:j2}
\end{equation}
The first piece---Eq. \ref{eq:j1}---constitutes an additional
magnetic moment and modifies the W-boson g-factor from its
Belinfante value of unity to its standard model value of 2. Using
\begin{equation}
{1\over
2}(\epsilon_{Bi}^*\epsilon_{Aj}+\epsilon_{Ai}\epsilon_{Bj}^*)-{1\over
3}\delta_{ij}\hat{\epsilon}_B^*\cdot\hat{\epsilon}_A=<1,m_f|{1\over
2}(S_iS_j+S_jS_i)-{2\over 3}\delta_{ij}|1,m_i>
\end{equation}
we observe that the second component---Eq. \ref{eq:j2}---implies
the existence of a quadrupole moment of size $Q=-e/M_W^2$. Both of
these results are well known predictions of the standard model for
the charged vector bosons\cite{pdg}.

In fact, it has recently been argued, from a number of viewpoints,
that the "natural" value of the gyromagnetic ratio for a particle
of {\it arbitrary} spin should be g=2\cite{gft}, as opposed to the
value $1/S$ from the Belinfante conjecture, and we shall
consequently employ g=2 in our spin 1 calculations below.

Having obtained the appropriate Langrangian for the interactions
of a charged spin-1 system,
\begin{equation}
{\cal L}(x)=-{1\over 2}U_{\mu\nu}^\dagger(x)
U^{\mu\nu}(x)+m^2\phi_\mu^\dagger(x)\phi^\mu(x)-eF^{\mu\nu}(x)
(\phi_\mu^{+\dagger}(x)\phi_\nu^+(x)-\phi_\mu^{-\dagger}(x)\phi^-_\mu(x))
\end{equation}
we can now calculate the matrix elements which will be needed in
our calculation.  Specifically, the general single photon vertex
for a transition involving an outgoing photon with polarization
index $\mu$ and four-momentum $q=p1-p2$, an incoming spin one
particle with polarization index $\alpha$ and four-momentum $p_1$
together with an outgoing spin one particle with polarization
index $\beta$ and four-momentum $p_2$ is\cite{bja}
\begin{eqnarray}
V^{(1)}_{\beta,\alpha,\mu}(p_1,p_2)=-ie\left[(p_1+p_2)_\mu\eta_{\alpha\beta}-
(gp_{1\beta}-(g-1)p_{2\beta})\eta_{\alpha\mu}-
(gp_{2\alpha}-(g-1)p_{1\alpha})\eta_{\beta\mu})\right]
\end{eqnarray}
while the two-photon vertex with polarization indices $\mu,\nu$,
an incoming spin one particle with polarization index $\alpha$ and
four-momentum $p_1$ together with an outgoing spin one particle
with polarization index $\beta$ and four-momentum $p_2$ has the
form\cite{bjb}
\begin{equation}
V^{(2)}_{\beta,\alpha,\mu\nu}(p_1,p_2)=-ie^2(2\eta_{\mu\nu}\eta_{\alpha\beta}
-\eta_{\alpha\mu}\eta_{\beta\nu}-\eta_{\alpha\nu}\eta_{\beta\mu})
\end{equation}
The energy-momentum tensor connecting an incoming vector meson
with polarization index $\alpha$ and four-momentum $k_1$ with and
outgoing vector with polarization index $\beta$ and four-momentum
$k_2$ is found to be\cite{jkn}
\begin{eqnarray}
<k_2,\beta|T_{\mu\nu}|k_1,\alpha>&=&(k_{1\mu}k_{2\nu}+k_{1\nu}k_{2\mu})\eta_{\alpha\beta}\nonumber\\
&-&k_{1\beta}(k_{2\mu}\eta_{\alpha\nu}+k_{2\nu}\eta_{\alpha\mu})\nonumber\\
&-&k_{2\alpha}(k_{1\nu}\eta_{\beta\mu}+k_{1\mu}\eta_{\beta\nu})\nonumber\\
&+&(k_1\cdot
k_2-m^2)(\eta_{\beta\mu}\eta_{\alpha\nu}+\eta_{\beta\nu}\eta_{\alpha\mu})\nonumber\\
&-&\eta_{\mu\nu}[(k_1\cdot
k_2-m^2)\eta_{\alpha\beta}-k_{1\beta}k_{2\alpha}]
\end{eqnarray}
and that between photon states is identical, except that the terms
in $m^2$ are absent.  For later use, it is useful to note that the
trace of this expression has the simple form
\begin{equation}
\eta^{\mu\nu}<k_2,\beta|T_{\mu\nu}|k_1,\alpha>=2m^2\eta_{\alpha\beta}
\end{equation}
which vanishes in the case of the photon.  The leading component
of the on-shell energy-momentum tensor between charged vector
meson states is then
\begin{eqnarray}
<k_2,\epsilon_B|T_{\mu\nu}^{(0)}|k_1,\epsilon_A>&=&(k_{1\mu}k_{2\nu}+k_{1\nu}k_{2\mu})
\epsilon_B^*\cdot\epsilon_A\nonumber\\
&-&k_1\cdot\epsilon_B^*(k_{2\mu}\epsilon_{A\nu}+k_{2\nu}\epsilon_{A\mu}\nonumber\\
&-&k_2\cdot\epsilon_A(k_{1\nu}\eta_{B\mu}^*+k_{1\mu}\epsilon_{B\nu}^*)\nonumber\\
&+&(k_1\cdot
k_2-m^2)(\epsilon_{B\mu}^*\epsilon_{A\nu}+\epsilon_{B\nu}^*\epsilon_{A\mu})\nonumber\\
&-&\eta_{\mu\nu}[(k1\cdot
k_2-m^2)\epsilon_B^*\cdot\epsilon_A-k_1\cdot\epsilon_B^*k_2\cdot\epsilon_A]\label{eqn:tm}
\end{eqnarray}
and the focus of our calculation is to evaluate the one-loop
electromagnetic corrections to Eq. \ref{eqn:tm}, via the diagrams
shown in Figure 1, keeping only the leading nonanalytic terms.
Details of the calculation are described in the appendix, and the
results are given by
\begin{itemize}
\item [a)] Seagull loop diagram (Figure 1a)
\begin{eqnarray}
Amp[a]_{\mu\nu}&=& {L\alpha\over 48\pi m}\left[\epsilon_A\cdot
\epsilon_B^*(2q_\mu q_\nu+{1\over 2
}q^2\eta_{\mu\nu})-\epsilon_A\cdot q \epsilon_B^*\cdot
q\eta_{\mu\nu}\right.\nonumber\\
&+&\left.\epsilon_A\cdot q(\epsilon_{B\mu}^*
q_\nu+\epsilon_{B\nu}^* q_\mu)+\epsilon_B^*\cdot q(\epsilon_{A\mu}
q_\nu+\epsilon_{A\nu} q_\mu)\right.\nonumber\\
&-&\left.2(\epsilon_{A\mu} \epsilon_{B\nu}^*+\epsilon_{A\nu}
\epsilon_{B\mu}^*)q^2\right]
\end{eqnarray}

\item [b)] Born loop diagram (Figure 1b)
\begin{eqnarray}
Amp[b]_{\mu\nu}&=&{\alpha\over 48\pi m}\left\{
-3P_\mu P_\nu q^2\epsilon_B^*\cdot\epsilon_A(8L+3S)\right.\nonumber\\
&+&\left.[(P_\mu
\epsilon_{A\nu}+P_\nu\epsilon_{A\mu})\epsilon_B^*\cdot q-(P_\mu
\epsilon_{B\nu}^*+P_\nu\epsilon_{B\mu}^*\epsilon_A\cdot
q)]q^2(4L+3S)\right.\nonumber\\
&-&\left.[\epsilon_A\cdot q(\epsilon_{B\mu}^*
q_\nu+\epsilon_{B\nu}^* q_\mu)+\epsilon_B^*\cdot q(\epsilon_{A\mu}
q_\nu+\epsilon_{A\nu}
q_\mu)]L\right.\nonumber\\
&-&\left.[q_\mu q_\nu(10L+3S)-q^2\eta_{\mu\nu}({15\over
2}L+3S)]\epsilon_B^*\cdot\epsilon_A\right.\nonumber\\
&+&\left.2(\epsilon_{B\mu}^*\epsilon_{A\nu}+\epsilon_{B\nu}^*\epsilon_{A\mu})q^2L\right.\nonumber\\
&+&\left.\ldots\right\}
\end{eqnarray}
\end{itemize}

Note that due to conservation of the energy-momentum
tensor---$\partial^\mu T_{\mu\nu}=0$---the on-shell matrix element
must satisfy the gauge invariance condition
$$q^\nu<k_2,\epsilon_B|T_{\mu\nu}|k_1,\epsilon_A>=0$$
In our case, although the leading order contribution satisfies
this condition
\begin{equation}
q^\mu<k_2,\epsilon_B|T_{\mu\nu}^{(0)}|k_1,\epsilon_A>=0
\end{equation}
the contribution from {\it neither} diagram 1a or 1b is
gauge-invariant
\begin{eqnarray}
q^\mu Amp[a]_{\mu\nu}&=&{\alpha L\over 96\pi m
}\left[5q^2q_\nu\epsilon_B^*\cdot\epsilon_A
-2q^2(\epsilon_{A\nu}\epsilon_B^*\cdot
q+\epsilon_{B\nu}^*\epsilon_A\cdot q))
+2q_\mu\epsilon_{B}^*\cdot q\epsilon_a\cdot q\right]\nonumber\\
q^\mu Amp[b]_{\mu\nu}&=&-{\alpha L\over 96\pi m
}\left[5q^2q_\nu\epsilon_B^*\cdot\epsilon_A
-2q^2(\epsilon_{A\nu}\epsilon_B^*\cdot
q+\epsilon_{B\nu}^*\epsilon_A\cdot q)) +2q_\mu\epsilon_{B}^*\cdot
q\epsilon_a\cdot q\right]\nonumber\\
\qquad
\end{eqnarray}
Nevertheless the sum of these terms {\it is} gauge-invariant---
\begin{equation}
q^\mu(Amp[a]_{\mu\nu}+Amp[b]_{\mu\nu})=0
\end{equation}
which serves as an important check on our result.

The full loop contribution is then
\begin{eqnarray}
Amp[a+b]_{\mu\nu}&=&{\alpha\over 48\pi
m}\left\{\epsilon_B^*\cdot\epsilon_A[(q_\mu
q_\nu-q^2\eta_{\mu\nu})-3P_\mu P_\nu
q^2](8L+3S)\right.\nonumber\\
&+&\left.((P_\mu
\epsilon_{A\nu}+P_\nu\epsilon_{A\mu})\epsilon_B^*\cdot q-(P_\mu
\epsilon_{B\nu}^*+P_\nu\epsilon_{B\mu}^*)\epsilon_A\cdot
q)q^2(4L+3S)\right.\nonumber\\
&+&\left.\ldots\right\}
\end{eqnarray}
Due to covariance and gauge invariance the form of the matrix
element of $T_{\mu\nu}$ between on-shell spin 1 states must be
expressible in the form
\begin{eqnarray}
&&<p_2,\epsilon_B|T_{\mu\nu}(x)|p_1,\epsilon_A>_{S=1}=-{e^{i(p_2-p_1)\cdot
x}\over \sqrt{4E_1E_2}}[2P_\mu P_\nu \epsilon_B^*\cdot
\epsilon_AF_1(^{(S=1)}q^2)\nonumber\\
&+&(q_\mu q_\nu-\eta_{\mu\nu}q^2)
\epsilon_B^*\cdot\epsilon_AF_2^{(S=1)}(q^2)\nonumber\\
&+&[P_\mu(\epsilon_{B\nu}^* \epsilon_A\cdot q-\epsilon_{A\nu}
\epsilon_B^*\cdot q)+P_\nu(\epsilon_{B\mu}^*
\epsilon_A\cdot q-\epsilon_{A\mu} \epsilon_B^*\cdot q)]F_3^{(S=1)}(q^2)\nonumber\\
&+&\left[(\epsilon_{A\mu}
\epsilon_{B\nu}^*+\epsilon_{B\mu}^*\epsilon_{A\nu})q^2-(\epsilon_{B\mu}^*
q_\nu+\epsilon_{B\nu}^* q_\mu)\epsilon_A\cdot q\right.\nonumber\\
&-&\left.(\epsilon_{A\mu} q_\nu+\epsilon_{A\nu}
q_\mu)\epsilon_B^*\cdot q+2\eta_{\mu\nu}\epsilon_A\cdot q
\epsilon_B^*\cdot q\right]F_4^{(S=1)}(q^2)\nonumber\\
&+&{2\over m^2}P_\mu P_\nu \epsilon_A\cdot q \epsilon_B^*\cdot q
F_5^{(S=1)}(q^2)\nonumber\\
&+&{1\over m^2}(q_\mu q_\nu-\eta_{\mu\nu}q^2)\epsilon_A\cdot
q\epsilon_B\cdot q F_6^{(S=1)}(q^2)]\label{eq:gi}
\end{eqnarray}
Here $F_1^{(S=1)}(q^2),F_2^{(S=1)}(q^2),F_3^{(S=1)}(q^2)$
correspond to their spin 1/2 counterparts $F_1^{(S={1\over 2}
)}(q^2),F_2^{(S={1\over 2})}(q^2),F_3^{(S={1\over 2})}(q^2)$,
while $F_4^{(S=1)}(q^2),F_5^{(S=1)}(q^2),F_6^{(S=1)}(q^2)$
represent new forms unique to spin 1.  (Note that each kinematic
form in Eq. \ref{eq:gi} is separately gauge invariant.)

The results of the calculation described above can most concisely
be described in terms of these form factors.  Thus the tree level
predictions can be described as
\begin{eqnarray}
F_{1,tree}^{(S=1)}&=&F_{3,tree}^{(S=1)}=1\nonumber\\
F_{2,tree}^{(S=1)}&=&F_{4,tree}^{(S=1)}=-{1\over 2}\nonumber\\
F_{5,tree}^{(S=1)}&=&F_{6,tree}^{(S=1)}=0
\end{eqnarray}
while the full radiatively corrected values are given by
\begin{eqnarray}
F_1^{(S=1)}(q^2)&=&1+{\alpha\over 16\pi}{q^2\over m^2}(8L+3S)+\ldots\nonumber\\
F_2^{(S=1)}(q^2)&=&-{1\over 2}+{\alpha\over 24\pi}(8L+3S)+\ldots\nonumber\\
F_3^{(S=1)}(q^2)&=&1+{\alpha\over 24\pi}{q^2\over m^2}(4L+3S)+\ldots\nonumber\\
F_4^{(S=1)}(q^2)&=&-{1\over 2}+{\alpha\over 192\pi}{q^2\over m^2}(16L-3S)+\ldots\nonumber\\
F_5^{(S=1)}(q^2)&=&{\alpha\over 384\pi}{q^2\over m^2}(64L+9S)+\ldots\nonumber\\
F_6^{(S=1)}(q^2)&=&{\alpha\over
192\pi}(64L+15S)+\ldots\label{eq:re}
\end{eqnarray}
and we note that $F_{1,2,3,loop}^{(S=1)}(q^2)$ as found for unit
spin agree exactly with the forms $F_{1,2,3,loop}^{(S={1\over
2})}(q^2)$found previously for spin 1/2 and with
$F_{1,2,loop}^{(S=0)}(q^2)$ in the spinless case.  We observe that
both $F_1^{(S=1)}(q^2=0)=1$ and $F_3^{(S=1)}(q^2=0)=1$ as required
by energy-momentum and angular momentum conservation.
Interestingly, the form factors $F_4^{(S=1)}(q^2)$ and
$F_5^{(S=1)}(q^2)$ are also unrenormalized from their tree level
values and this fact has an interesting consequence.  Since, in
the Breit frame and using nonrelativistic kinematics we have
\begin{eqnarray}
&&<p_2,\epsilon_B|T_{00}(0)|p_1,\epsilon_A>\simeq
m\left\{\hat{\epsilon}_B^*\cdot\hat{\epsilon}_AF_1^{(S=1)}(q^2)
+{1\over
2m^2}\hat{\epsilon}_B^*\cdot\vec{q}\hat{\epsilon}_A\cdot\vec{q}\right.\nonumber\\
&\times&\left.
[F_1^{(S=1)}(q^2)-F_2^{(S=1)}(q^2)-2(F_4^{(S=1)}(q^2)+F_5^{(S=1)}(q^2)
-{q^2\over 2 m^2}F_6^{(S=1)}(q^2))]
\right\}+\ldots\nonumber\\
&&<p_2,\epsilon_B|T_{0i}(0)|p_1,\epsilon_A>\simeq -{1\over
2}[(\hat{\epsilon}_B^*\times\hat{\epsilon}_A)\times\vec{q}]_iF_3^{(S-1)}(q^2)+\ldots
\end{eqnarray}
we can identify values for the gravitoelectric monopole,
gravitomagnetic dipole, and gravitoelectric quadrupole coupling
constants
\begin{eqnarray}
K_{E0}&=&mF_1^{(S=1)}(q^2=0)\nonumber\\
K_{M1}&=&{1\over
2}F_3^{(S=1)}(q^2=0)\nonumber\\
K_{E2}&=&{1\over
2m}\left[F_1^{(S=1)}(q^2=0)-F_3^{(S=1)}(q^2=0)-2F_4^{(S=1)}(q^2=0)-2F_5^{(S=1)}(q^2=0)\right]\nonumber\\
\end{eqnarray}
Taking $Q_g\equiv m$ as the gravitational "charge," we observe
that the tree level values---
\begin{equation}
K_{E0}=Q_g\qquad K_{M1}={Q_g\over 2m}\qquad K_{E2}={Q_g\over m^2}
\end{equation}
are {\it unrenormalized} by loop corrections.  That is to say, not
only does there not exist any anomalous gravitomagnetic moment, as
mentioned above, but also there is no anomalous gravitoelectric
quadrupole moment.

As an aside, we note that an additional check on our results is
found by using the feature pointed out above that the radiative
correction terms must be traceless.  In our case this means that
two conditions must be satisfied, since there are two separate
scalar structures---$\epsilon_A\cdot\epsilon_B$ and
$\epsilon_A\cdot q\epsilon_B\cdot q$---whose coefficients must
vanish, which leads to
\begin{eqnarray}
\epsilon_A\cdot\epsilon_B&:&P^2F_1^{(S=1)}(q^2)-{3\over
2}{q^2\over
m^2}F_2^{(S=1)}(q^2)+2q^2F_4^{(S=1)}(q^2)=0\nonumber\\
\epsilon_A\cdot q\epsilon_B\cdot q
&:&F_3^{(S=1)}(q^2)+4F_4^{(S=1)}(q^2)+{P^2\over m^2
}F_5^{(S=1)}(q^2)-3{q^2\over m^2}F_6^{(S=1)}(q^2)=0\nonumber\\
\quad
\end{eqnarray}
Both strictures are satisfied at the level to which we work.

\subsection{An Addendum}

Before concluding, it is interesting to note that although we have
used the standard model value $g=2$ which is appropriate for a
charged W-boson, it is not necessary to do so.  Indeed for a
charged spin 1 system like the $\rho^\pm$ strong interaction
corrections lead to a very different value of the gyromagnetic
ratio, and it is interesting to calculate the form factors which
would result from an arbitrary choice of $g$.  The results are
found to be
\begin{eqnarray}
F_1^{(S=1)}(q^2)&=&1+{\alpha\over 16\pi}{q^2\over m^2}(8L+3S)+\ldots\nonumber\\
F_2^{(S=1)}(q^2)&=&-{1\over 2}+{\alpha\over 24\pi}(8L+3S)+\ldots\nonumber\\
F_3^{(S=1)}(q^2)&=&1+{\alpha\over 48\pi}{q^2\over m^2}((-(4+g(g-8))L+3gS)+\ldots\nonumber\\
F_4^{(S=1)}(q^2)&=&-{1\over 2}+{\alpha\over 1536\pi}{q^2\over m^2}(16(g(3g-4)+4)L-3(g(g+4)-4)S)+\ldots\nonumber\\
F_5^{(S=1)}(q^2)&=&-{\alpha\over 3072\pi}{q^2\over m^2}(32(g-6)(g+2)L+3(g(5g-12)-20)S)+\ldots\nonumber\\
F_6^{(S=1)}(q^2)&=&{\alpha\over
1536\pi}(32(g+2)^2L+3(g(20-3g)+12)S)+\ldots\label{eq:gm}
\end{eqnarray}
Of course, one easily confirms that in the limit $g\rightarrow 2$
these results reproduce those given in Eq. \ref{eq:re}.  However,
there is something else of interest here.  In ref. \cite{gar} it
was shown that the classical piece of the leading form factors
follows simply from the requirement that the long range component
of the matrix element of the energy-momentum tensor agree with the
well known form
\begin{equation}
T_{\mu\nu}=-F_{\mu\lambda}{F_\nu}^\lambda+{1\over
4}g_{\mu\nu}F_{\alpha\beta}F^{\alpha\beta}\label{eq:em}
\end{equation}
For the diagonal components of $T_{\mu\nu}$, this leads for a
point charge $e$ to
\begin{eqnarray}
T_{00}&=&{1\over 2}\vec{E}^2={\alpha\over 8\pi r^4}\nonumber\\
T_{0i}&=&0\nonumber\\
 T_{ij}&=&-E_iE_j+{1\over
2}\delta_{ij}\vec{E}^2=-{\alpha\over 4\pi r^4}\left({r_ir_j\over
r^2}-{1\over 2}\delta_{ij}\right)
\end{eqnarray}
which can easily be shown to correspond to Fourier transform of
the leading diagonal classical loop corrections to the
energy-momentum tensor, {\it i.e.,} those terms of the loop
calculation proportional to S---
\begin{eqnarray}
T_{00}^{loop,cl}(\vec{r})&=&\int {d^3k\over
(2\pi)^3}e^{-i\vec{k}\cdot\vec{r}}T_{00}^{loop,cl}(\vec{k})={\alpha\over 8\pi r^4}\nonumber\\
T_{0i}^{loop,cl}(\vec{r})&=&0\nonumber\\
T_{ij}^{loop,cl}(\vec{r})&=&\int{d^3k\over
(2\pi)^3}e^{-i\vec{k}\cdot\vec{r}}T_{ij}^{loop,cl}(\vec{k})=-{\alpha\over
4\pi r^4}\left({r_ir_j\over r^2}-{1\over 2}\delta_{ij}\right)
\end{eqnarray}
where we have used the forms
\begin{eqnarray}
T_{00}^{loop,cl}(\vec{k})&=&-{\alpha\pi|\vec{k}|\over 8}\nonumber\\
T_{ij}^{loop,cl}(\vec{k})&=&{\alpha\pi\over
16|\vec{k}|}(k_ik_j-\delta_{ij}\vec{k}^2)
\end{eqnarray}
found above.

However, when the charged particle carries spin, a magnetic field
\begin{equation}
\vec{B}={ge\over 2m}{3\hat{r}\vec{S}\cdot\hat{r}-\vec{S}\over 4\pi
r^3}
\end{equation}
must also be included.  This changes the diagonal pieces of the
energy-momentum tensor by calculable higher order terms, which
means that the {\it leading} diagonal pieces determined by the
form factors $F_1(q^2),F_2(q^2)$ should be unchanged and therefore
independent of $g$ as found in Eq. \ref{eq:gm}.  On the other hand
the presence of a magnetic field leads to a nonvanishing
off-diagonal component
\begin{equation}
T_{0i}=-(\vec{E}\times\vec{B})_i=-{\alpha g\over 8\pi
mr^6}(\vec{S}\times\vec{r})_i
\end{equation}
which means that the classical component of our loop correction to
the off-diagonal component of the energy momentum tensor must also
have this form
\begin{equation}
T_{0i}^{loop,cl}(\vec{r})=\int {d^3k\over
(2\pi)^3}e^{-i\vec{k}\cdot\vec{r}}T_{0i}^{loop,cl}(\vec{k})
\end{equation}
From Eq. \ref{eq:gm} we find
\begin{equation}
T_{0i}^{loop,cl}(\vec{k})=-i{\alpha\pi g|\vec{k}|\over 32
m}(\vec{S}\times\vec{k})_i
\end{equation}
whose Fourier transform yields the expected form.  Thus the linear
proportionality of the classical loop correction to $F_3(q^2)$ to
the gyromagnetic ratio $g$ is expected, and it is only if the
value $g=2$ is employed that the Kerr-Newman metric is reproduced.
Nevertheless, we find that other consistent values to the metric
tensor are reproduced if alternative values of the gyromagnetic
ratio apply.  In the case of the quantum component of the loop
correction the ratio of the spin 1 and spin 1/2 terms is found to
be
\begin{equation}
{F_3^{loop,qm}\over F_3^{loop,qm}}=g(1-{g\over 8})-{1\over 2}
\end{equation}
so that again use of the value $g=2$ guarantees universality for
both the classical and the quantum correction as well as brings
these results into conformity with the Kerr-Newman form.  This
point deserves further study.

\section{Conclusion}

Above we have calculated the radiative corrections to the
energy-momentum tensor of a spin 1 system.  We have confirmed the
universality which was speculated in our previous work in that we
have confirmed that
\begin{eqnarray}
F_{1,loop}^{(S=0)}(q^2)&=&F_{1,loop}^{(S={1\over
2})}(q^2)=F_{1,loop}^{(S=1)}(q^2)\nonumber\\
F_{2,loop}^{(S=0)}(q^2)&=&F_{2,loop}^{(S={1\over
2})}(q^2)=F_{2,loop}^{(S=1)}(q^2)\nonumber\\
F_{3,loop}^{(S={1\over 2})}(q^2)&=&F_{3,loop}^{(S=1)}(q^2)
\end{eqnarray}
The universality in the case of the classical (square root)
nonanalyticities is not surprising and in fact is {\it required}
by the connection to the metric tensor and to the classical form
of the energy-momentum tensor---Eqn. \ref{eq:em}.  In the case of
the quantum (logarithmic) nonanalyticities, however, it is not
clear why these terms must be universal.  We also found additional
higher order form factors for spin 1, which also receive loop
corrections. It is tempting to conclude that this radiative
correction universality holds for arbitrary spin. However, it is
probably not possible to show this by generalizing the
calculations above. Indeed the spin 1 result involves {\it
considerably} more computation than does its spin 1/2 counterpart,
which was already much more tedious than that for spin 0.  Perhaps
a generalization such as that used in nuclear beta decay can be
employed\cite{nbd}. Work is underway on this problem and will be
reported in an upcoming communication.

\begin{center}
{\bf Acknowledgement}
\end{center}

This work was supported in part by the National Science Foundation
under award PHY-02-42801.  Useful conversations with John Donoghue
and Andreas Ross are gratefully acknowledged, as is the
hospitality of Prof. A. Faessler and the theoretical physics group
from the University of T\"{u}bingen, where this paper was
finished.

\section{Appendix}

In this section we sketch how our results were obtained.  The
basic idea is to calculate the Feynman diagrams shown in Figure 1.
Thus for Figure 1a we find\cite{ppr}
\begin{equation}
Amp[a]_{\mu\nu}={1\over 2!}\int{d^4k\over
(2\pi)^4}{\epsilon_B^{*\beta}V^{(2)}_{\beta,\alpha,\mu\nu}
(p1,p2)\epsilon_A^\alpha <k-q,\beta|
T_{\mu\nu}|k,\alpha>|_{m^2=0}\over k^2(k-q)^2}\label{eqn:a}
\end{equation}
while for Figure 1b\cite{ppr}
\begin{eqnarray}
Amp[b]_{\mu\nu}&=&\int{d^4k\over (2\pi)^4}{1\over
k^2(k-q)^2((k-p)^2-m^2)}\nonumber\\
&\times&\epsilon_B^\beta
V^{(1)}_{\beta,\lambda,\kappa}(p_2,p_1-k)\left(-\eta^{\lambda\zeta}+{(p1-k)^\lambda
(p_1-k)^\zeta\over
m^2}\right)\nonumber\\
&\times&V^{(1)}_{\zeta,\rho,\delta}(p1-k,p1)\epsilon_A^\rho
<k,\kappa|T_{\mu\nu}|k-q,\delta>_{m^2=0}\label{eqn:b}
\end{eqnarray}
Here the various vertex functions are listed in section 3, while
for the integrals, all that is needed is the leading nonanalytic
behavior.  Thus we use
\begin{eqnarray}
I(q)&=&\int{d^4k\over (2\pi)^4}{1\over k^2(k-q)^2}={-i\over
32\pi^2}(2L+\ldots)\nonumber\\
I_\mu(q)&=&\int{d^4k\over (2\pi)^4}{k_\mu\over k^2(k-q)^2}={i\over
32\pi^2}(q_\mu L+\ldots)\nonumber\\
I_{\mu\nu}(q)&=&\int{d^4k\over (2\pi)^4}{k_\mu k_\nu\over
k^2(k-q)^2}={-i\over 32\pi^2}(q_\mu q_\nu{2\over
3}L-q^2\eta_{\mu\nu}{1\over 6}L +\ldots)\nonumber\\
I_{\mu\nu\alpha}(q)&=&\int{d^4k\over (2\pi)^4}{k_\mu k_\nu
k_\alpha\over k^2(k-q)^2}={i\over 32\pi^2}(-q_\mu q_\nu q_\alpha
{L\over 2}\nonumber\\
&+&(\eta_{\mu\nu}q_\alpha+\eta_{\mu\alpha}q_\nu
+\eta_{\nu\alpha}q_\mu){1\over 12}Lq^2 +\ldots)\nonumber\\
\quad
\end{eqnarray}
for the "bubble" integrals and
\begin{eqnarray}
J(p,q)&=&\int{d^4k\over (2\pi)^4}{1\over
k^2(k-q)^2((k-p)^2-m^2)}={-i\over
32\pi^2m^2}(L+S)+\ldots\nonumber\\
J_\mu(p,q)&=&\int{d^4k\over (2\pi)^4}{k_\mu\over
k^2(k-q)^2((k-p)^2-m^2)}={i\over
32\pi^2m^2}\nonumber\\
&\times&[p_\mu((1+{1\over 2}{q^2\over m^2})L-{1\over 4}{q^2\over
m^2}S)-q_\mu(L+{1\over
2}S)+\ldots]\nonumber\\
J_{\mu\nu}(p,q)&=&\int{d^4k\over (2\pi)^4}{k_\mu k_\nu\over
k^2(k-q)^2((k-p)^2-m^2)}={i\over 32\pi^2m^2}\nonumber\\
&\times&[-q_\mu q_\nu(L+{3\over 8}S)-p_\mu p_\nu{q^2\over
m^2}({1\over 2}L+{1\over 8}S)\nonumber\\
&+&q^2\eta_{\mu\nu}({1\over 4}L+{1\over 8}S)+(q_\mu p_\nu+q_\nu
p_\mu)(({1\over 2}+{1\over 2}{q^2\over m^2})L+{3\over 16}{q^2\over
m^2 S})\nonumber\\
J_{\mu\nu\alpha}(p,q)&=&\displaystyle\int\frac{d^4k}{(2\pi)^4}
\frac{k_\mu k_\nu k_\alpha}{k^2(k-q)^2((k-p)^2-m^2)} \nonumber\\
&=& \frac{-i}{32\pi^2m^2}\bigg[ q_\mu q_\nu
q_\alpha\bigg(L+\frac5{16}S\bigg)+p_\mu p_\nu
p_\alpha\bigg(-\frac16 \frac{q^2}{m^2}\bigg) \nonumber\\
\nonumber&+&\big(q_\mu p_\nu p_\alpha + q_\nu p_\mu p_\alpha +
q_\alpha p_\mu p_\nu\big)\bigg(\frac13\frac{q^2}{m^2}L+
\frac1{16}\frac{q^2}{m^2}S\bigg)\nonumber\\&+&\big(q_\mu q_\nu
p_\alpha + q_\mu q_\alpha p_\nu + q_\nu q_\alpha p_\mu
\big)\bigg(\Big(-\frac13 - \frac12\frac{q^2}{m^2}\Big)L
-\frac{5}{32}\frac{q^2}{m^2}S\bigg)\nonumber\\
\nonumber &+&\big(\eta_{\mu\nu}p_\alpha + \eta_{\mu\alpha}p_\nu +
\eta_{\nu\alpha}p_\mu\big)\Big(\frac1{12}q^2L\Big)\nonumber\\
\nonumber&+&\big(\eta_{\mu\nu}q_\alpha + \eta_{\mu\alpha}q_\nu +
\eta_{\nu\alpha}q_\mu\big)\Big(-\frac16q^2L -\frac1{16}q^2S\Big)
\bigg]+\ldots\nonumber\\
\quad
\end{eqnarray}
for their "triangle" counterparts.  Similarly higher order forms
can be found, either by direct calculation or by requiring various
identities which must be satisfied when the integrals are
contracted with $p^\mu,q^\mu$ or with $\eta^{\mu\nu}$.  Using
these integral forms and substituting into Eqs. \ref{eqn:a} and
\ref{eqn:b}, one derives the results given in section 3.


\begin{thebibliography}{99}
\bibitem{fta} The Fourier transforms can be found in Appendix C of
\cite{gar}.
\bibitem{bjb} This form agrees with that given by Bjorken and
Drell in their appendix.
\bibitem{jkn} The $m^2$-independent piece of this matrix element
can be found from the simple energy-momentum tensor for the
electromagnetic field
$$T_{\mu\nu}^{em}=-F_{\mu\lambda}{F_\nu}^\lambda+
{1\over 4}\eta_{\mu\nu}F_{\lambda\delta}F^{\lambda\delta}$$
\end{thebibliography}
\end{document}